\documentclass[11pt, oneside]{article}   	% use "amsart" instead of "article" for AMSLaTeX format
\usepackage{amsmath, amsthm, amssymb}
\usepackage{tabu}
\usepackage{caption}
\usepackage{array}
\usepackage{geometry}                		% See geometry.pdf to learn the layout options. There are lots.
\usepackage{mathtools}
\usepackage{morefloats}
\usepackage[parfill]{parskip}    		% Activate to begin paragraphs with an empty line rather than an indent
\usepackage{graphicx}				% Use pdf, png, jpg, or eps with pdflatex; use eps in DVI mode
\usepackage{caption}								% TeX will automatically convert eps --> pdf in pdflatex	
\usepackage{textcomp}
\usepackage{cite}
\usepackage{gensymb}

\usepackage{enumitem}												
\usepackage{amssymb}

\title{Intermediate Mirrors to Reach Theoretical Efficiency Limits of Multi-Bandgap Solar Cells}
\author{Vidya Ganapati, Chi-Sing Ho, Eli Yablonovitch\\ \\University of California, Berkeley}
\begin{document}
\maketitle
\thispagestyle{empty}
\pagestyle{empty}
%\section{}
%\subsection{}

%$$ $$ - separate line equation
%$ $

%{} \frac {}

\section*{Abstract}

Creating a single bandgap solar cell that approaches the Shockley-Queisser limit requires a highly reflective rear mirror. This mirror enhances the voltage of the solar cell by providing photons with multiple opportunities for escaping out the front surface. Efficient external luminescence is a pre-requisite for high voltage. Intermediate mirrors in a multijunction solar cell can enhance the voltage for each cell in the stack. These intermediate mirrors need to have the added function of transmitting the below bandgap photons to the next cell in the stack. In this work, we quantitatively establish the efficiency increase possible with the use of intermediate selective reflectors between cells in a tandem stack. The absolute efficiency increase can be up to $\approx 6\%$ in dual bandgap cells with optimal intermediate and rear mirrors. A practical implementation of an intermediate selective mirror is an air gap sandwiched by antireflection coatings. The air gap provides perfect reflection for angles outside the escape cone, and the antireflection coating transmits angles inside the escape cone. As the incoming sunlight is within the escape cone, it is transmitted on to the next cell, while most of the internally trapped luminescence is reflected.

\section*{Introduction} 

In the last few years, the efficiency record for a single bandgap solar cell has risen to $28.8\%$ \cite{green_solar_2014}, a record held by a thin-film gallium arsenide cell from Alta Devices. This efficiency increase was enabled by improvements in the optical design \cite{miller_strong_2012, miller_photonic_2012}. This record-holding single bandgap cell had a rear reflector, rather than a substrate. In a solar cell, some of the absorbed photons will be radiatively emitted in the material. These internally luminescent photons can then be re-absorbed in the cell (and then either lost to non-radiative recombination or radiatively emitted again) or escape from a surface. In an ideal solar cell with perfect rear mirror, all the internally luminescent photons will eventually escape from the top surface. A good rear reflector provides multiple opportunities for a luminescent photon to escape out of the front surface of the cell, and was instrumental in achieving the record single bandgap solar cell efficiency \cite{miller_strong_2012}. 

In a multijunction solar cell, bandgaps of different materials are placed in a stack, from largest bandgap on top to smallest on the bottom. The top cell absorbs all the photons above its bandgap, and the lower energy photons are transmitted to the next bandgap. In the past year, a new record of $31.1\%$ was set by the National Renewable Energy Laboratory, for a dual bandgap solar cell under 1 sun illumination, by taking advantage of the improved voltage in the bottom cell \cite{nrel_2013}. In a multijunction solar cell stack, improving the rear reflector improves the voltage of the bottommost cell, however the upper cells do not get this same voltage boost. In order to further improve the efficiency of a dual bandgap solar cell, an intermediate mirror needs to be placed in between the top and bottom cells. This intermediate mirror needs to reflect the internally luminescent photons (which are mostly at the bandgap energy, arriving at all angles), and to transmit the sub-bandgap photons to the cell below. These sub-bandgap photons have a range of energy, but are near normal incidence, owing to the refraction into the higher index solar cell material. 

\section*{Limiting Efficiency of Multijunction Cells}

The quasi-equilibrium derivation given by Shockley and Queisser \cite{shockley_detailed_1961} yields the limiting efficiency of a solar cell with one material bandgap. References \cite{brown_detailed_2002,henry_limiting_2008,marti_limiting_1996,parrott_limiting_1979,tobias_ideal_2002,vos_detailed_1980} extend the analysis to multiple bandgaps, obtaining the limiting efficiencies with multiple material bandgaps. Of these, \cite{brown_detailed_2002,henry_limiting_2008,tobias_ideal_2002} analyze the case where the cells are electrically connected in series, so each cell must operate at the same current. Nonetheless, in our following theoretical analysis of the multijunction cell, we assume that each cell is electrically independent (i.e. each cell has two terminal connections), in order to find limiting efficiencies. References \cite{marti_limiting_1996,tobias_ideal_2002,vos_detailed_1980} look at the case where there are no intermediate mirrors and all the cells are index matched. Multijunction cells with perfect intermediate mirrors (defined here as a mirror which reflects all above bandgap photons and transmits all below bandgap photons) were analyzed in \cite{brown_detailed_2002,marti_limiting_1996}, but the effect of improved luminescence extraction in boosting the voltage was not accounted for. Here, we account for the voltage boost that arises from improved external extraction from each bandgap of a tandem cell. 

We derive the limiting efficiency of multijunction cells following a similar procedure to the derivation for single bandgap cells in \cite{miller_strong_2012}. We assume step function absorption (all photons above the bandgap energy are absorbed, and all photons below the bandgap energy are transmitted).

We will first consider the top cell, which consists of the material with the largest bandgap, $E_{g1}$. The analysis of this top cell is identical to the single bandgap case derived in \cite{miller_strong_2012}. The analysis begins in the dark, at thermal equilibrium, with the cell absorbing blackbody radiation from the external environment. The blackbody radiation $b(E)$ can be approximated by the tail of the blackbody formula:

\begin{equation}\label{eq:1bb} b(E)=\frac{2E^2} {h^3 c^2} exp\left( -\frac{E} {kT} \right),\end{equation}

where the units of $b$ are [photons/(time $\times$ area $\times$ energy $\times$ steradian)]. $E$ is the photon energy, $h$ is Planck's constant, $c$ is the speed of light, and $kT$ is the thermal energy. 
The photon flux through the front surface of the solar cell due to absorption of the blackbody is given as:

\begin{equation}\label{eq:2absBB} 2 \pi \int \int_0^{\frac{\pi}{2}} a(E,\theta) b(E) \sin\theta \cos\theta d\theta dE, \end{equation}

where $\theta$ is the angle from the normal to the cell, and $a(E,\theta) = a(E)$ is the step function absorptivity for $E_{g1}$. Since the cell is in thermal equilibrium, this expression is also equivalent to the photon flux emitted out of the front surface. When the sun illuminates the cell, it moves into quasi-equilibrium, with chemical potential $qV$ (this is equivalent to the separation of the quasi-Fermi levels, where $q$ is the charge of an electron and $V$ is the voltage). Under illumination, the photon flux out the front of the cell, $L_{ext}$, is given by:

\begin{equation}\label{eq:3Lext} L_{ext}=exp\left(\frac{qV}{kT}\right) 2 \pi \int \int_0^{\frac{\pi}{2}} a(E,\theta) b(E) \sin\theta \cos\theta d\theta dE. \end{equation}

 On the front surface of the cell is a perfect antireflection coating. We assume that the solar cell is in air with index of refraction $n = 1$. Thus, at the top surface, we can assume perfect transmittance of internally luminescent photons in the escape cone $\theta_s$  (given by Snell's law, $n_s \sin \theta_s = 1$, where $n_s$ is the refractive index of the top semiconductor). There is total internal reflection for internal luminescent photons outside the escape cone. We assume that the internal luminescence hitting the top surface has a Lambertian distribution, due to the strong absorption of the material (the material is optically thick to the luminescent photons here; we will later consider the case of an optically thin material). The angle averaged transmittance of the internally luminescent photons through the top surface, $T_{int}$, is thus given by:

\begin{equation}\label{eq:4Tint} T_{int}=\frac{2 \pi \int_0^{sin^{-1}\left(\frac{1}{n_s}\right)} \sin\theta \cos\theta d\theta} {2 \pi \int_0^{\frac{\pi}{2}} \sin\theta \cos\theta d\theta}=\frac{1}{n_s^2}.\end{equation}

We assume the cell is free of non-radiative recombination in this analysis. The only other photon flux out of the cell is the flux out of the rear of the cell, which is described by rear luminescent transmittance $T_{int\downarrow}$. The external luminescence yield, $\eta_{ext}$, is defined as the ratio of the rate of radiative flux out the top, $L_{ext}$, to the total emission rate of photons $L_{ext}+L_{int\downarrow}$, where $L_{int\downarrow}$ is the radiative flux out the bottom to the cell below:

\begin{equation}\label{eq:5next} \eta_{ext}=\frac{L_{ext}}{L_{ext}+L_{int\downarrow} }=\frac{T_{int}}{T_{int}+T_{int\downarrow}}.\end{equation}

The absorption of photons from the sun is $\int a(E)S(E) dE$, where $S$ is the number of photons in the solar spectrum per unit area per unit time. The current of the solar cell is given by the absorption of photons from the sun minus the emission of photons out of the cell. From Eqn.~\ref{eq:5next}, we get $L_{ext}+L_{int\downarrow}=\frac{L_{ext}}{\eta_{ext}}$. Thus the $JV$ characteristic of the solar cell is given by:

\begin{equation}\label{eq:6JV1} J = \int_{E_{g1}}^{\infty} S(E) dE - L_{ext}-L_{int\downarrow}=  \int_{E_{g1}}^{\infty} S(E) dE - \frac{1}{\eta_{ext}} \, \pi \, exp\left(\frac{qV_1}{kT}\right)\int_{E_{g1}}^{\infty} b(E) dE, \end{equation}

where $J$ is the current density and $V_1$ is the voltage of the top cell. The $JV$ curve of the top cell in the tandem stack is given by Eqn.~\ref{eq:6JV1}. The value of $V_1$ should be chosen to be the maximum power point of the cell.

The expression for the open circuit voltage of the top cell is given by setting $J=0$ in Eqn.~\ref{eq:6JV1}:

\begin{equation}\label{eq:7Voc} V_{oc,1} = \frac{kT}{q} ln\left(\frac{ \int_{E_{g1}}^{\infty} S(E) dE}{\pi \int_{E_{g1}}^{\infty} b(E) dE}\right) - \frac{kT}{q} ln\left(\frac{1}{\eta_{ext}}\right). \end{equation}

From Eqn.~\ref{eq:7Voc}, we see that the open circuit voltage penalty when $\eta_{ext}<1$ is $\frac{kT}{q} ln\left(\frac{1}{\eta_{ext}}\right)$.

We now consider the second cell beneath the first cell. The absorption of photons from the sun is now given as $ \int_{E_{g2}}^{E_{g1}}S(E) dE$, (assuming step function absorptivity for the second cell as well). In the $JV$ characteristic of the second cell, there is an extra term to account for the radiative flux out of the bottom of the top cell that is absorbed by the second cell. Since from Eqn.~\ref{eq:5next}, $L_{int\downarrow}=\frac{L_{ext}}{\eta_{ext}} -L_{ext}$, the downward flux is given by:

\begin{equation}\label{eq:8rad} \left(\frac{1}{\eta_{ext}}-1\right) \pi \, exp\left(\frac{qV_1}{kT}\right)\int_{E_{g1}}^{\infty} b(E) dE. \end{equation}

By analogy to Eqn.~\ref{eq:6JV1} the $JV$ characteristic of the second cell is thus given by:

\begin{equation}\label{eq:9JV2} J = \int_{E_{g2}}^{E_{g1}} S(E) dE + \left(\frac{1}{\eta_{ext,1} }-1\right) \pi \, exp\left(\frac{qV_1}{kT}\right)\int_{E_{g1}}^{\infty} b(E) dE - \frac{1}{\eta_{ext,2}} \, \pi \, exp\left(\frac{qV_2}{kT}\right)\int_{E_{g2}}^{\infty} b(E) dE, \end{equation}

where $\eta_{ext,1}$ refers to the external fluorescence yield of the top cell, and $\eta_{ext,2}$ refers to the second cell. The derivation of the $JV$ characteristic for cells below the second follows the same procedure.

\section*{Comparison of Different Intermediate Reflectors}

\subsection*{Case (1): No Intermediate Mirror}

\begin{figure}[htbp]
\begin{center}
\includegraphics[scale=.5]{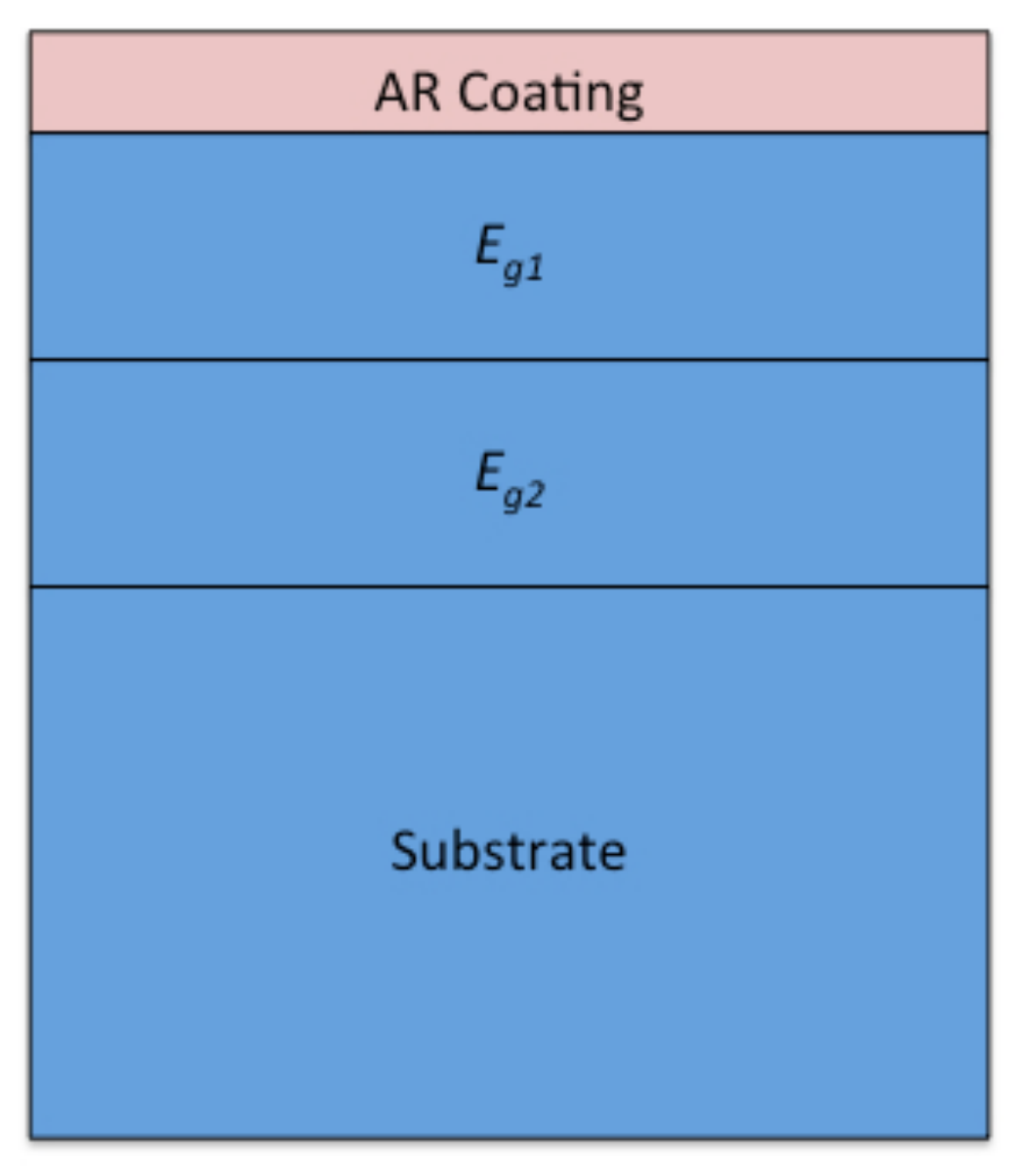}
\caption{Case (1); a dual bandgap solar cell without an intermediate or a back mirror; the top cell, bottom cell, and substrate are index matched with $n_s=3.5$. In Case (1a), the cells are optically thin to the luminescent photons, in Case (1b), the cells are optically thick.}
\label{Case1}
\end{center}
\end{figure}

\begin{figure}[htbp]
\begin{center}
\includegraphics[scale=1]{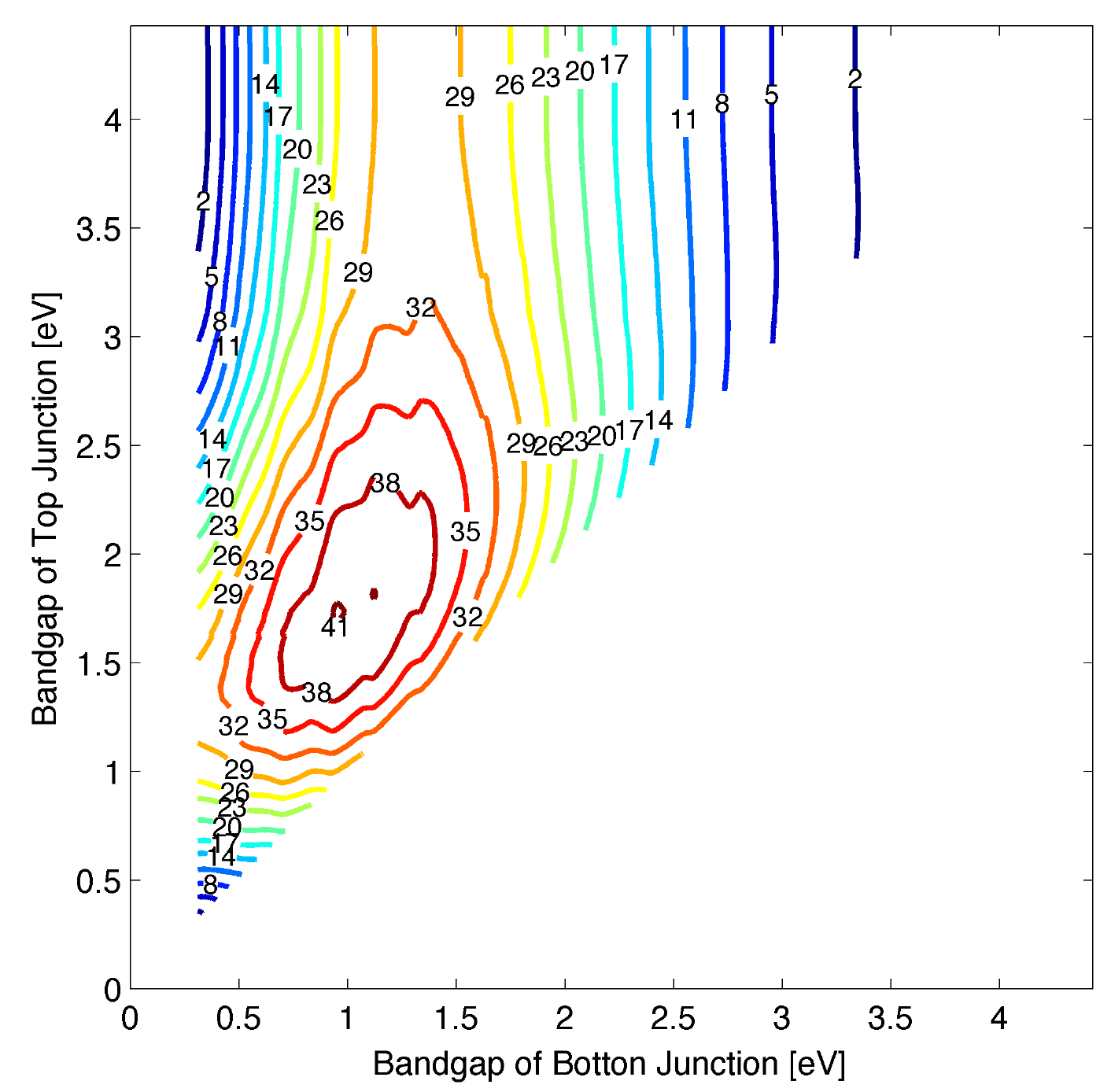}
\caption{The efficiencies as a function of top and bottom bandgap for Case (1a), a dual junction solar cell without an intermediate or a back mirror; the cells are assumed to be optically thin to the internally luminescent photons.}
\label{1aContours}
\end{center}
\end{figure}

We first consider the case of a dual bandgap solar cell without an intermediate mirror or rear mirror (see Fig.~\ref{Case1}). The top and bottom cells are index matched, on an absorbing substrate, and we assume a perfect antireflection coating on the top cell. In Eqns.~\ref{eq:4Tint} and \ref{eq:5next}, we assume a Lambertian distribution of internally luminescent photons, as the material is assumed to be optically thick to these photon energies. In Case (1a), we assume that the cells are optically thin to the luminescent photon energies. The external luminescence yield, $\eta_{ext}$, can also be described as the probability that an absorbed photon escapes out the front surface \cite{miller_strong_2012}. For the limit of a very optically thin cell, we can determine that $ \eta_{ext} \approx \frac{1}{4n_s^2}$ by recognizing that the probability of front surface escape, relative to substrate absorption, is the fraction of solid angle that is subtended by the escape cone \cite{yablonovitch_statistical_1982}. We plot the efficiencies as a function of bandgap in Fig.~\ref{1aContours}, assuming that the cells are optically thin to internal luminescence with no intermediate or rear mirror. In Fig.~\ref{1aContours}, and in the following calculations, we assume cell temperature of $T = 30 \degree C$, two terminal connections to each cell, 1 sun concentration, and an index of refraction of $n_s = 3.5$ for all the cells. We model the radiation from the sun with the Air Mass 1.5 Global tilt spectrum \cite{am15}.

\begin{figure}[htbp]
\begin{center}
\includegraphics[scale=1]{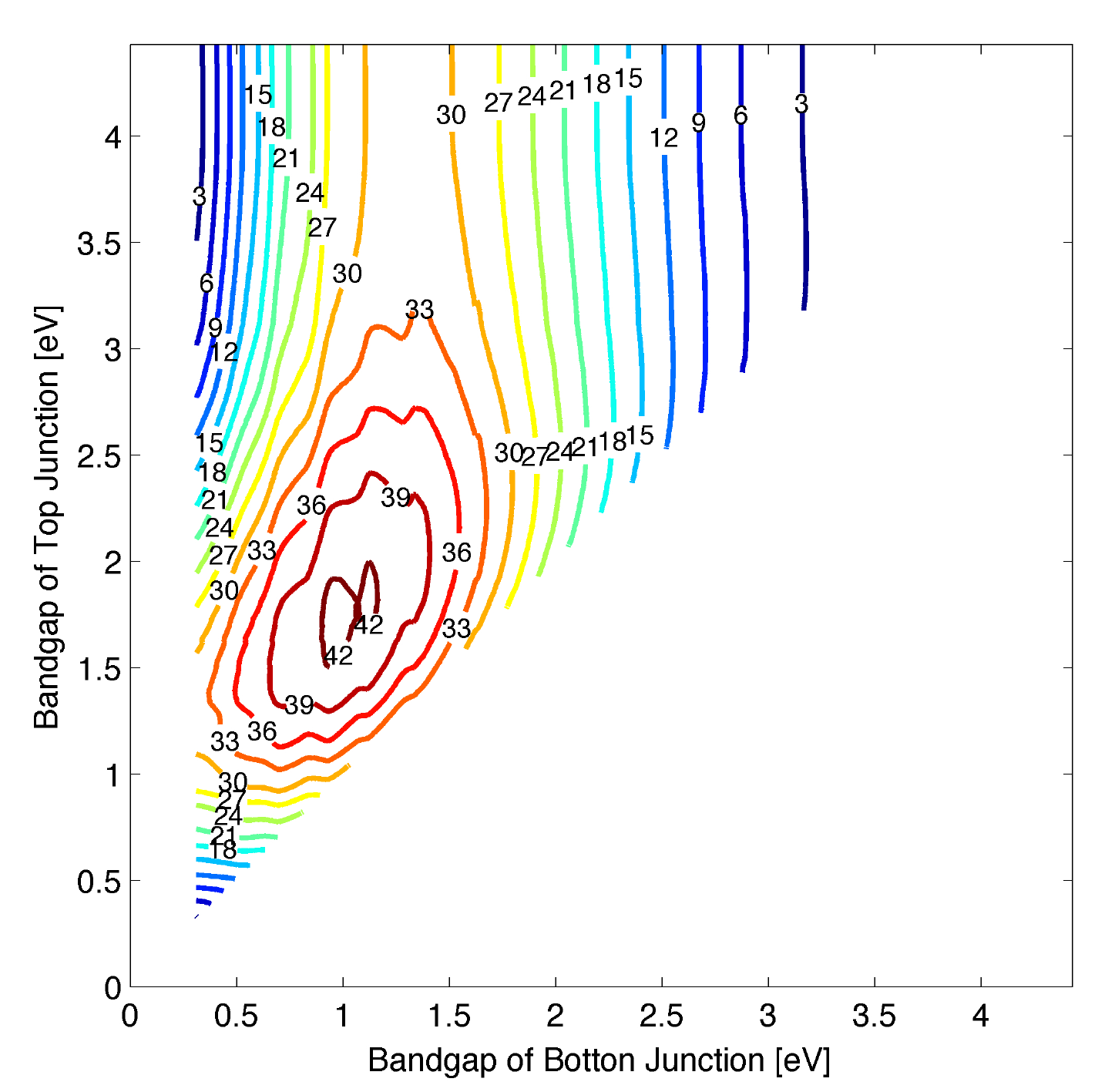}
\caption{The efficiencies as a function of top and bottom bandgap for Case (1b), a dual junction solar cell without an intermediate or a back mirror; the cells are assumed to be optically thick to the internally luminescent photons.}
\label{1bContours}
\end{center}
\end{figure}

In Case (1b), we assume that the cells are optically thick. We can then use Eqns.~\ref{eq:4Tint} and \ref{eq:5next} to determine $\eta_{ext}=\frac{1}{1+n_s^2}$ . We plot the efficiencies for this case in Fig.~\ref{1bContours}. We have a factor of $\approx \frac{1}{4}$ difference in $\eta_{ext}$ between the cases of optically thin and thick. We can account for this factor as follows: if the cell is optically thin, both the upward and downward luminescence are lost in the absorbing substrate, providing the first factor of 1/2. Furthermore, grazing incidence radiation will escape out the back, and not be diminished by $\langle \cos \theta \rangle = 0.5$ as in the Lambertian case, penalizing $\eta_{ext}$ with another factor of 1/2.  Thus, we obtain $\eta_{ext} \approx \frac{1}{4n_s^2}$  for optically thin, as opposed to $\eta_{ext} \approx \frac{1}{n_s^2}$  for optically thick. The impact of absorption on $\eta_{ext}$ is also discussed in \cite{steiner_optical_2013}.

\subsection*{Case (2): An Air Gap Intermediate Mirror}

An intermediate mirror for a dual bandgap cell must satisfy the requirements of (1) reflecting the internally luminescent photons of the top cell and (2) transmitting the externally incident photons that are below the bandgap of the top cell but above the bandgap of the bottom cell.

These dual requirements for an intermediate mirror appear difficult to satisfy, as we must satisfy them for photons at all energies and angles. Air gaps provide the following opportunity:
\begin{enumerate}[label={(\arabic*)}]
\item We obtain total internal reflection for the photons outside of the escape cone, as described by Snell's law. Due to the high index mismatch between the semiconductor and air, most of the internally luminescent photons are outside the escape cone and are thus reflected.
\item The externally incident photons, upon entrance into our structure, refract into the escape cone of the top cell material, as described by Snell's law. Thus, we can use antireflection coatings to transmit the photons in the escape cone to the next cell.
\end{enumerate}
The internally luminescent photons are created at all angles, while the transmitted solar photons have a limited angular range. Therefore angular filtering by an air gap can be employed instead of spectral filtering, to recycle the luminescent photons. 

\begin{figure}[htbp]
\begin{center}
\includegraphics[scale=.5]{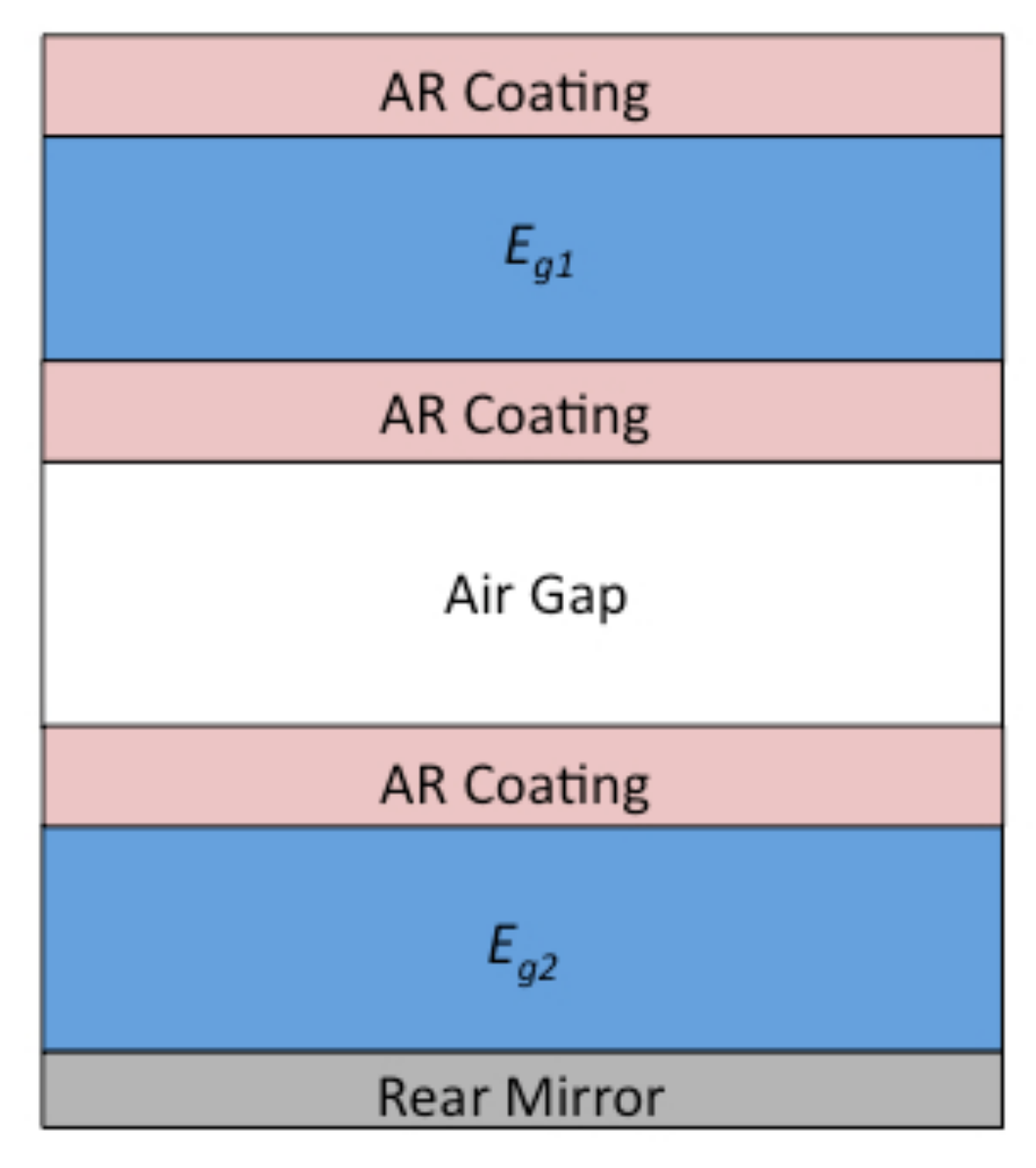}
\caption{Case (2); a dual bandgap solar cell with an air gap intermediate mirror, with perfect antireflection (AR) coatings and a perfect rear mirror.}
\label{Case2}
\end{center}
\end{figure}

We assume an air gap for the intermediate mirror, sandwiched by perfect antireflection coatings, as well as a perfect rear mirror and perfect top antireflection coating, see Fig.~\ref{Case2}. In this scenario, $\eta_{ext,1}=0.5$, as the front and back interfaces of the top cell are identical. With a perfect back reflector, $\eta_{ext,2}=1$, as all the photons must eventually escape out the front of the device. In this case, we obtain the same results whether the cells are optically thick or optically thin to the internal luminescence. For the top cell, the interfaces are symmetric, so optical thickness does not matter. For the bottom cell, a perfect mirror means that $\eta_{ext}=1$, regardless of optical thickness.

\begin{figure}[htbp]
\begin{center}
\includegraphics[scale=1]{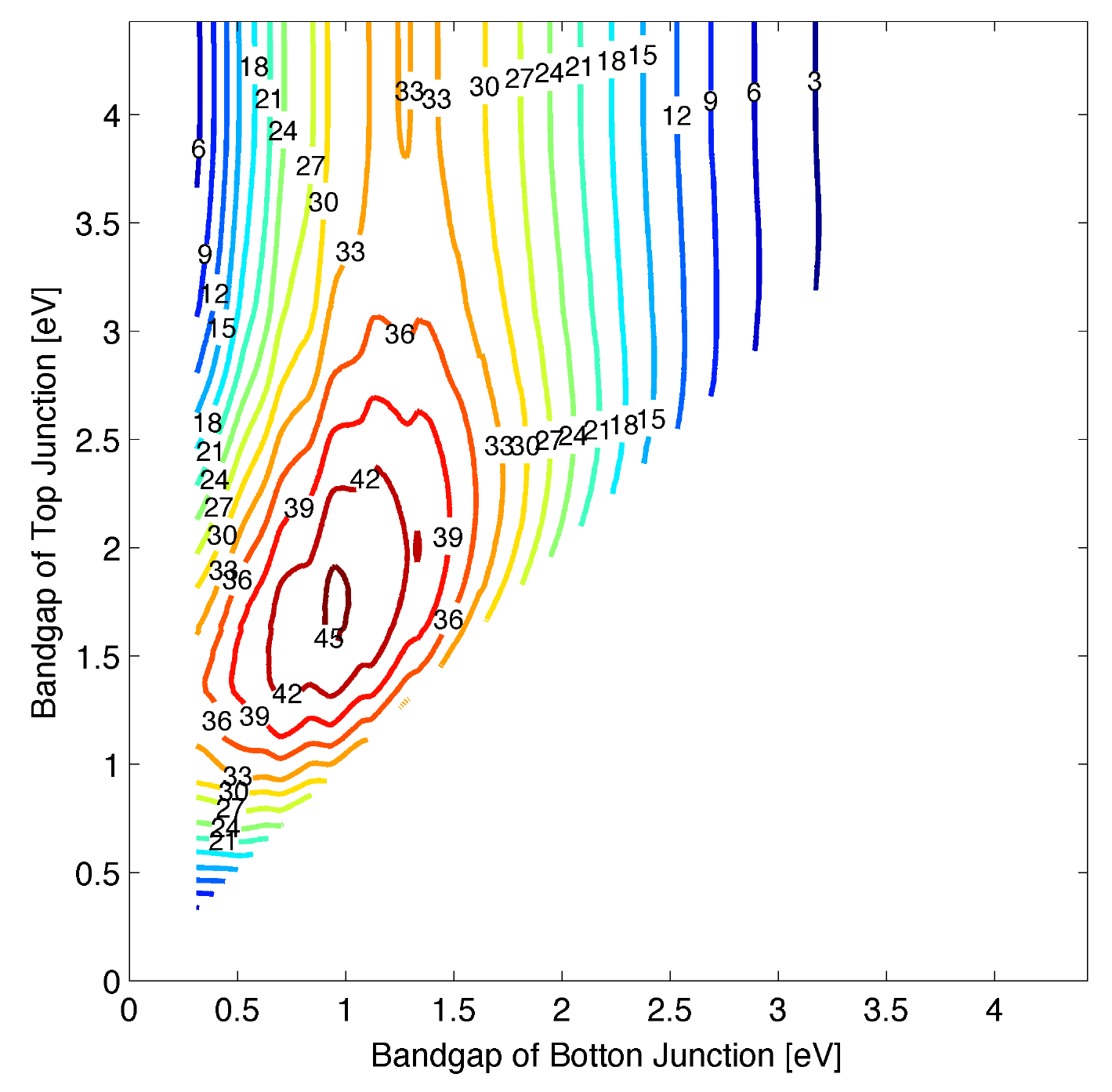}
\caption{The efficiencies as a function of top and bottom bandgap for Case (2), a dual junction cell with an air gap intermediate mirror and perfect rear mirror.}
\label{Case2Contours}
\end{center}
\end{figure}

\begin{figure}[htbp]
\begin{center}
\includegraphics[scale=1]{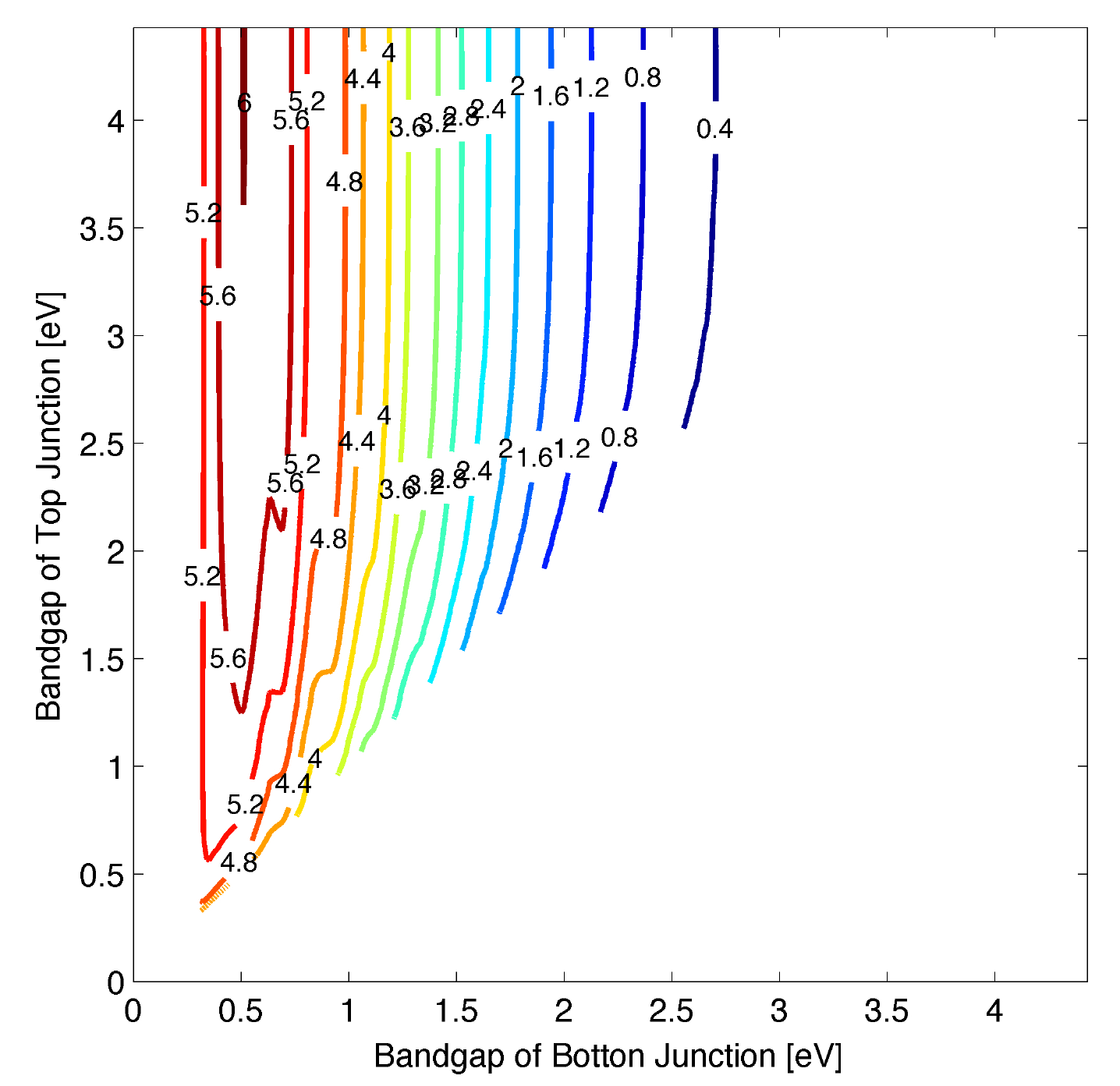}
\caption{Case (2) minus Case (1a); The absolute efficiency difference of cells with an air gap intermediate mirror and perfect rear mirror (see Fig.~\ref{Case2Contours}), and no mirrors, assuming optically thin cells (see Fig.~\ref{1aContours}). Up to $Å6\%$ can be picked up with an air gap intermediate mirror and perfect rear mirror if the cells are optically thin to the internal luminescence.}
\label{Diff2and1a}
\end{center}
\end{figure}

In Fig.~\ref{Case2Contours}, we plot the efficiency of the dual bandgap cell as a function of top and bottom bandgaps, assuming an air gap intermediate mirror, sandwiched by perfect antireflection coatings, as well as a perfect rear mirror, and perfect top antireflection coating. In Fig.~\ref{Diff2and1a}, we plot the difference in absolute efficiency between this case of an air gap intermediate mirror, and Case (1a), the case of no mirrors with cells assumed to be optically thin (as in Fig.~\ref{1aContours}). We can pick up $>6\%$ absolute in efficiency from the inclusion of an air gap and a rear mirror, for some pairs of bandgaps.

\subsection*{Case (3): Perfect Intermediate Mirror}

\begin{figure}[htbp]
\begin{center}
\includegraphics[scale=.5]{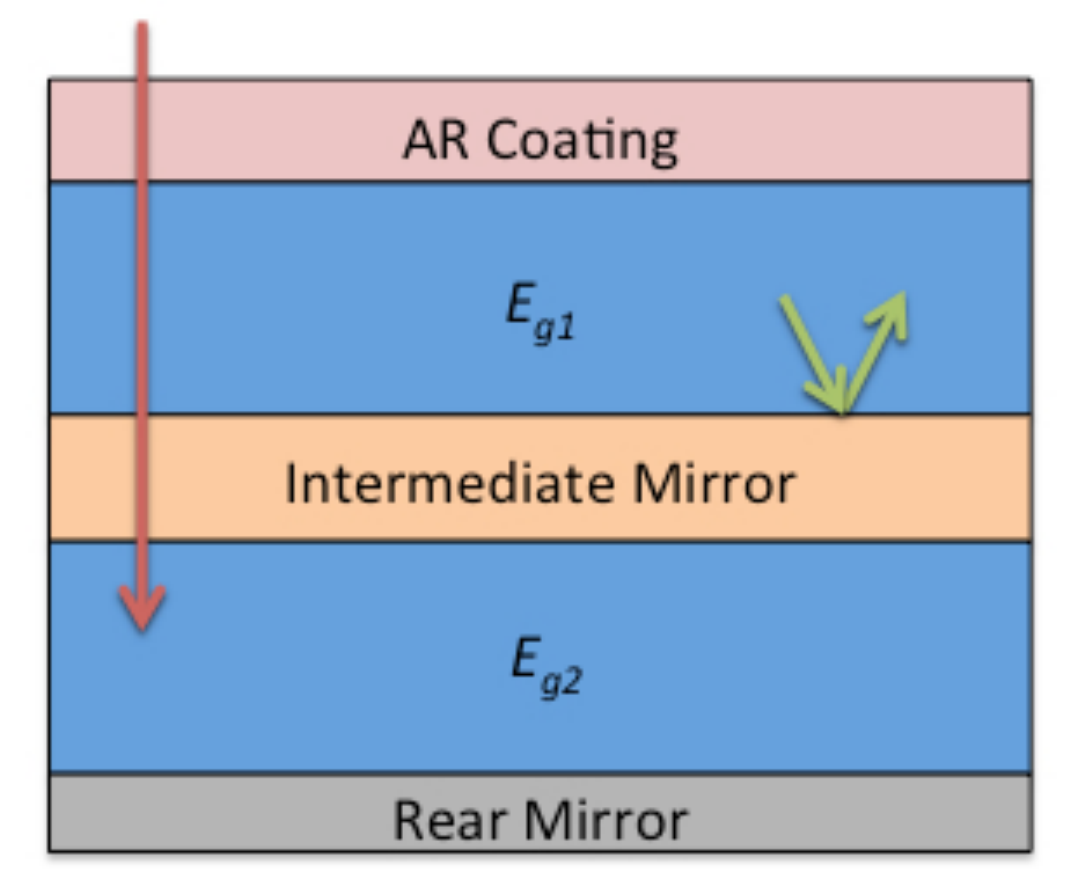}
\caption{Case (3); A dual bandgap solar cell with a perfect intermediate mirror and perfect rear mirror. A perfect intermediate mirror in a dual junction design must transmit photons with energy $< E_{g1}$, and reflect photons at energy $> E_{g1}$.}
\label{Case3}
\end{center}
\end{figure}

\begin{figure}[htbp]
\begin{center}
\includegraphics[scale=1]{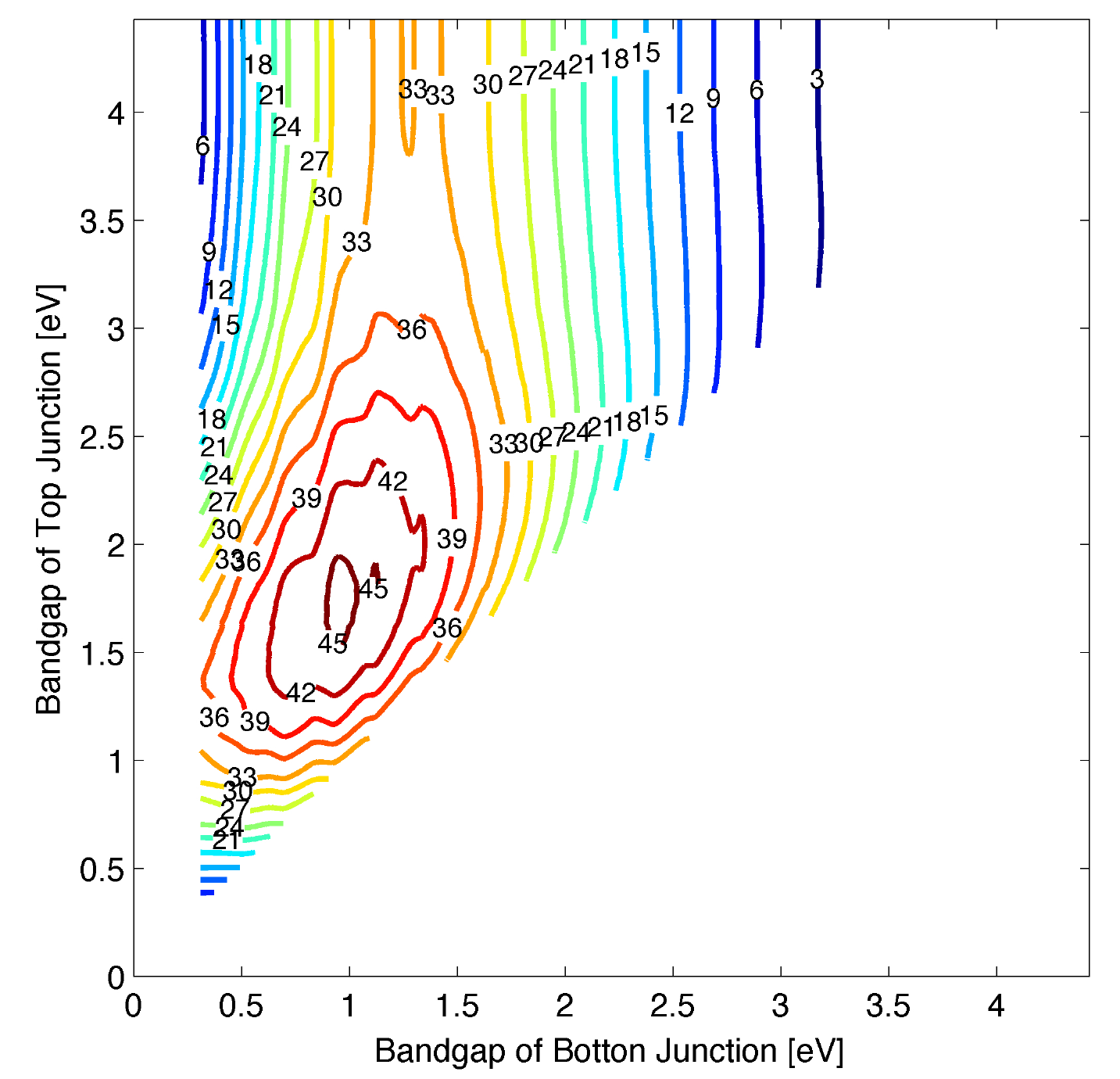}
\caption{Case (3); The theoretical efficiency limit of dual junction solar cells, as a function of top and bottom bandgap, assuming perfect intermediate and back mirrors.}
\label{Case3Contours}
\end{center}
\end{figure}

\begin{figure}[htbp]
\begin{center}
\includegraphics[scale=1]{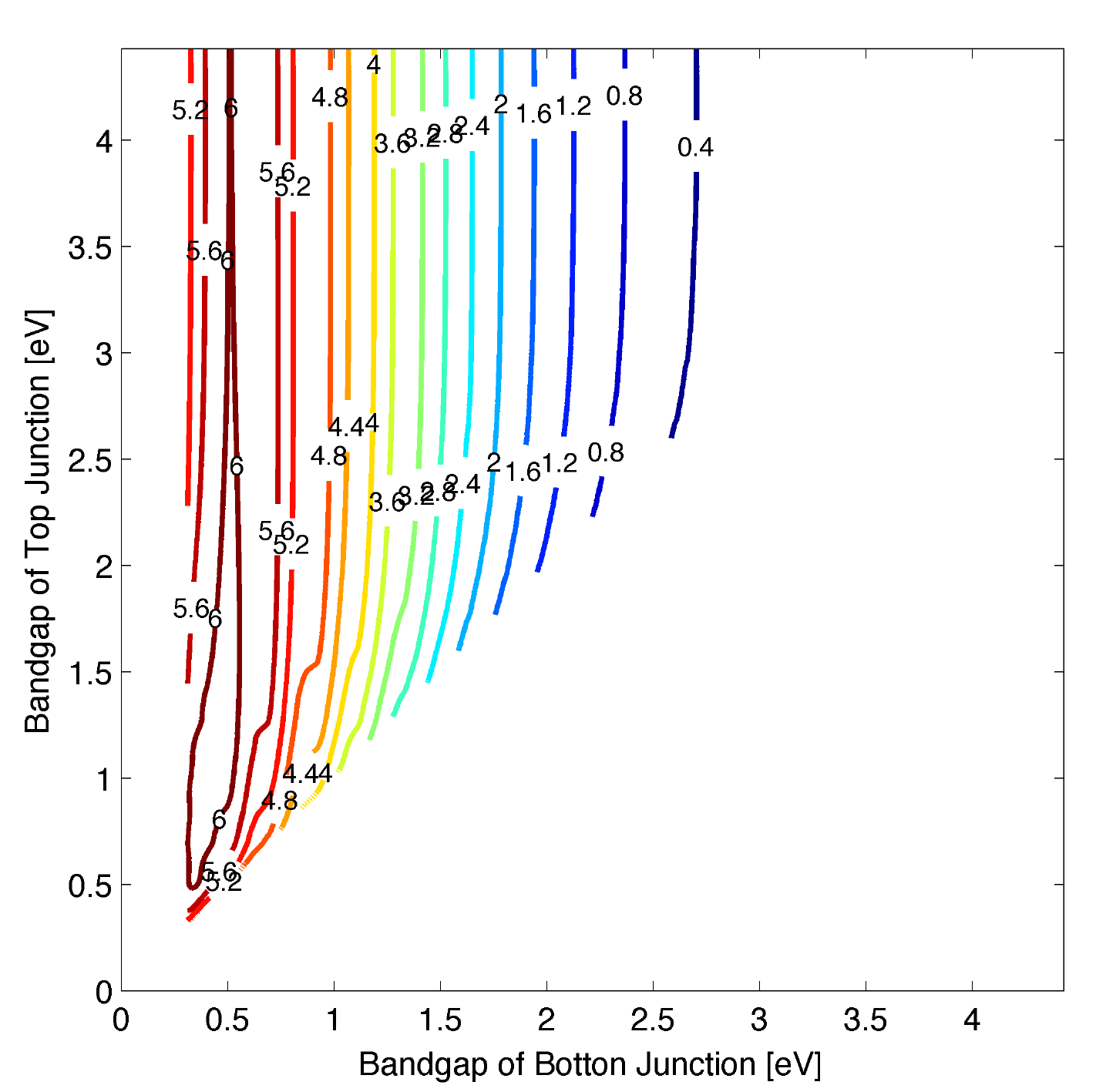}
\caption{Case (3) minus Case (1a); The absolute efficiency difference of cells with perfect mirrors (see Fig. 8), and no mirrors, assuming optically thin cells (see Fig. 2). Up to $Å6\%$ can be picked up with perfect mirrors if the cells are optically thin to the internal luminescence.}
\label{Diff3and1a}
\end{center}
\end{figure}

We finally consider the ideal case of a perfect intermediate mirror, perfect rear mirror, and perfect top antireflection coating; see Fig.~\ref{Case3}. The perfect intermediate mirror reflects all the photons above the bandgap of $E_{g1}$ and transmits all the photons below the bandgap of $E_{g1}$. For this case, $\eta_{ext,1} =\eta_{ext,2}=1$. We plot the limiting efficiencies for a dual bandgap cell with these ideal conditions in Fig.~\ref{Case3Contours}. Fig.~\ref{Diff3and1a} shows the absolute efficiency difference between this case and Case (1a).

\section*{Optimal 2-Bandgap Cell}

\begin{figure}[htbp]
\begin{center}
\includegraphics[scale=.5]{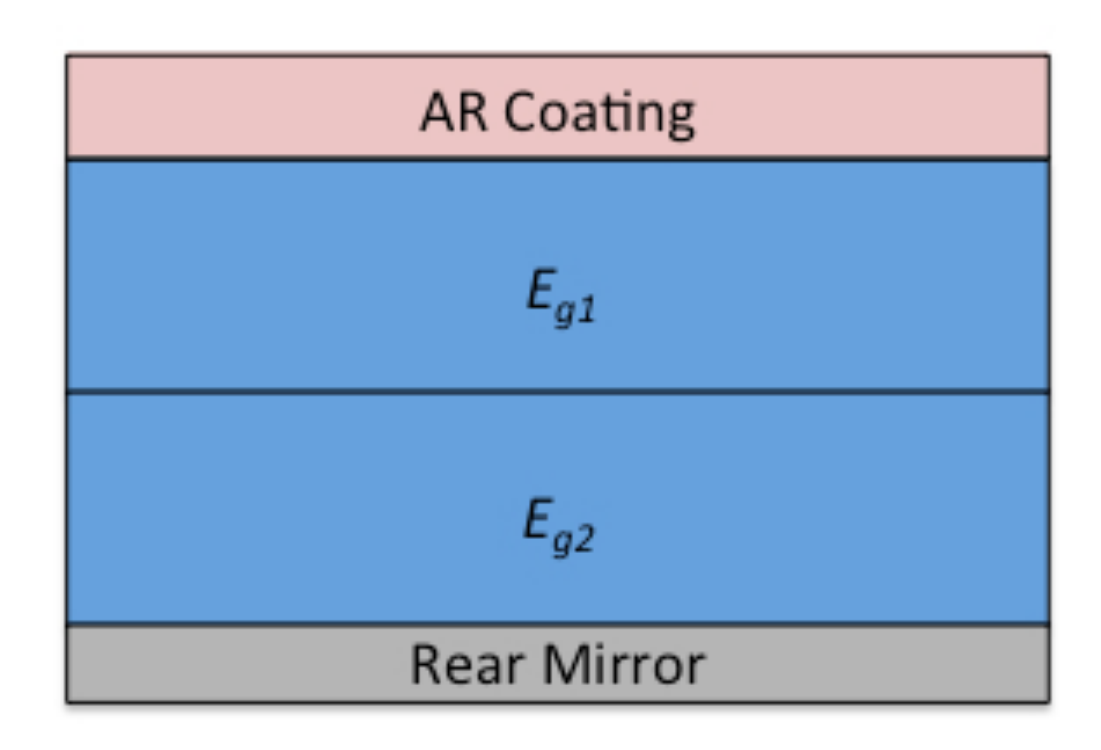}
\caption{Case (1c); A dual bandgap solar cell with no intermediate mirror and a perfect rear mirror. The top cell is assumed to be optically thin to the internal luminescence. Photons re-emitting out the back of the top cell are absorbed by the bottom cell. An intermediate mirror would both increase the voltage of the top cell and decrease the current of the bottom cell. In Case (1d), we have the same setup, except we assume the top cell is optically thick to the internal luminescence.}
\label{Case1c}
\end{center}
\end{figure}

\begin{figure}[htbp]
\begin{center}
\includegraphics[scale=.6]{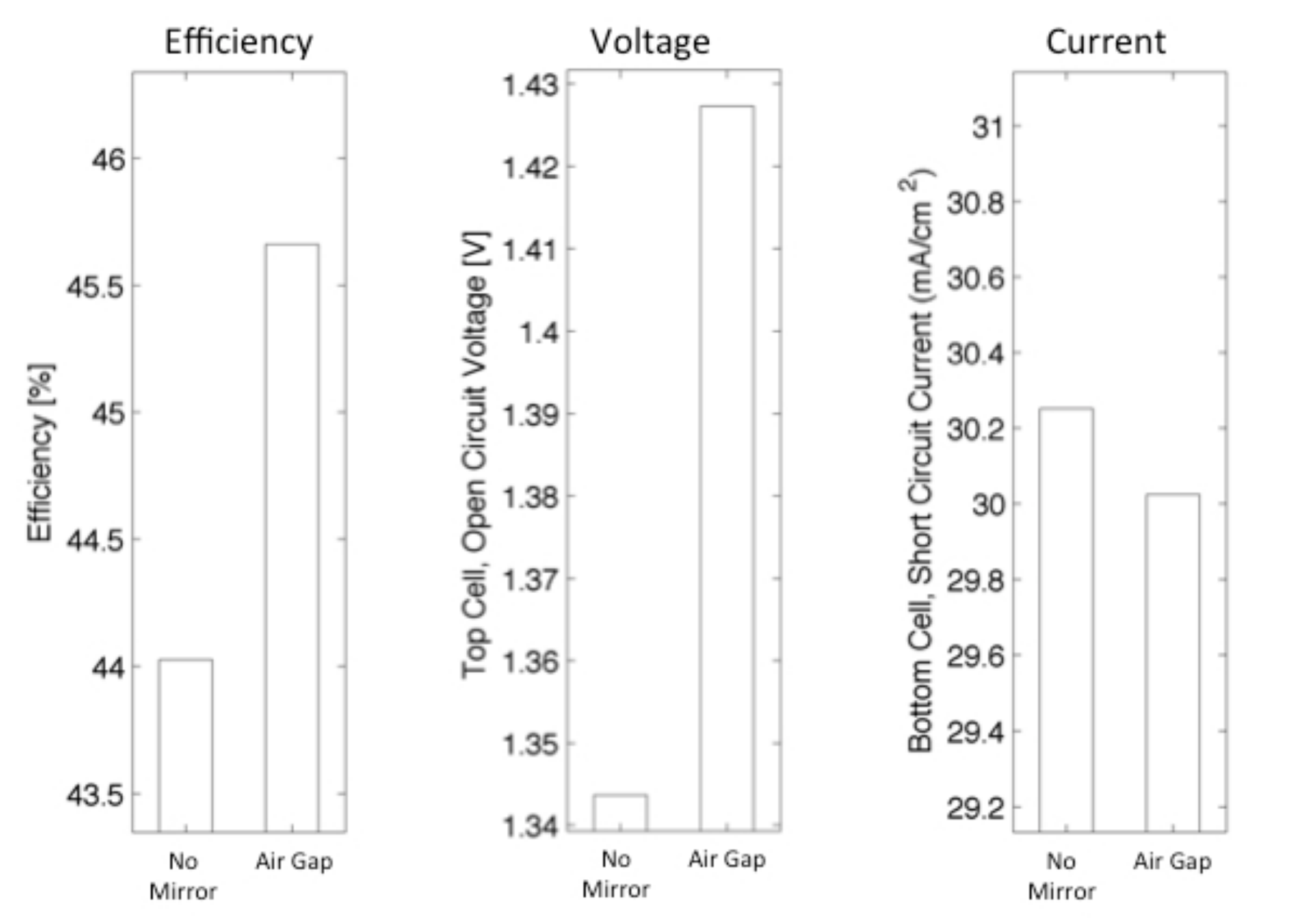}
\caption{Tandem cell efficiency, top cell open circuit voltage, and bottom cell current for bandgaps $E_{g1}$ = 1.73 eV and $E_{g2}$ = 0.94 eV, for Case (1c); no intermediate mirror and a perfect back mirror, assuming optically thin cells and Case (2); an air gap intermediate mirror and a perfect back mirror.}
\label{BarGraph}
\end{center}
\end{figure}

To isolate the effect of the intermediate mirror, we look at Case (1c), a dual bandgap solar cell with no intermediate mirror and a perfect rear mirror; see Fig.~\ref{Case1c}. The top cell is assumed to be optically thin to the internal luminescence. In Fig.~\ref{BarGraph}, we plot the open circuit voltage of the top cell, short circuit current of the bottom cell, and overall cell efficiency for Case (1c) and Case (2), the case with the air gap intermediate mirror and perfect rear mirror. The optimal bandgaps from Case (3), the ideal case, are used in this calculation ($E_{g1}$=1.73 eV and $E_{g2}$=0.94 eV). 

Eqn.~\ref{eq:7Voc} allows us to calculate the open circuit voltage penalties from the ideal Case (3). The thermal voltage is 26 mV, so in Case (2), with the air gap intermediate mirror, the top cell sees a voltage drop of $26 \, mV \times ln(2) =18 \, mV$. With no intermediate mirror, and an optically thin top cell as in Case (1c), the top cell sees a voltage drop of $26 \, mV  \times ln\left(4n_s^2\right)=100 \, mV$, with $n_s=3.5$. Thus, as we see in Fig.~\ref{BarGraph}, the top cell voltage difference between Case (2) and Case (1c) is $\approx 70$ mV .

As a result of the intermediate mirror, there is also a slight decrease in current in the bottom cell. This current decrease is due to the loss of radiative emission out the rear of the top cell that is then absorbed by the bottom cell. The effect of current loss in the bottom cell is not enough to offset the effect of gain in voltage of the top cell with the intermediate mirror. Thus for Case (2) minus case (1c), the tandem efficiency increases by $\approx 2\%$ with the air gap intermediate mirror.

\section*{Multijunction Cells}

Using the methodology described in the previous section, we calculate the limiting efficiency of multi-bandgap cells with 1 through 6 bandgaps, for Cases (1a), (1b), (1c), (2), and (3); see Table I. We also include the efficiency for Case (1d), which is similar to Case (1c) with no intermediate mirror and a perfect back mirror, except in Case (1d) we assume that the cells are optically thick to the internal luminescence; see Fig.~\ref{Case1c}.

\begin{figure}[htbp]
\begin{center}
\includegraphics[scale=.6]{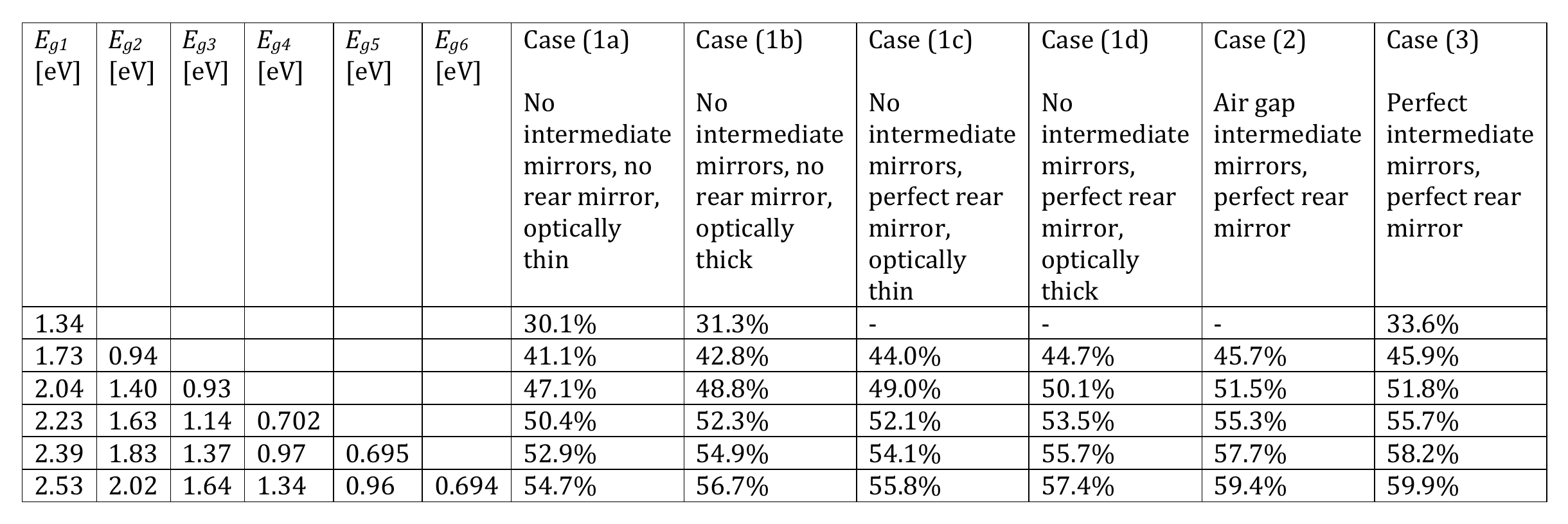}
\caption*{Table I: Efficiencies for cells with 1 through 6 bandgaps, for Cases (1a), (1b), (1c), (1d), (2), and (3). The efficiencies are calculated at the optimal bandgaps. The bandgaps for 4-6 cells are taken from \cite{brown_detailed_2002}.}
\label{Table1}
\end{center}
\end{figure}

In this paper, we have assumed $n_s=3.5$ for the refractive indices of all the cells. Due to the large refractive index mismatch with air ($n=1$), the escape cone given by Snell's law is $\sin^{-1}\left(\frac{1}{3.5}\right)  \approx 17\degree$ from the normal. Thus, when there is no rear mirror, the photon escape probability from the top surface, $\eta_{ext}$, is greatly diminished. 

We see a boost of $\approx 1\%$ absolute by adding an intermediate mirror for cells that are optically thick to the luminescence; Case (3) minus Case (1d) for a dual bandgap cell. In the work by Mart\'{i} and Ara\'{u}jo \cite{marti_limiting_1996}, the efficiency boost they calculate from adding an intermediate mirror in a similar situation is only $\approx 0.2\%$ absolute. This is because there is no refractive index mismatch with air in their work, and thus the luminescent photons have a very easy time escaping, so a mirror to assist the luminescence escape is unnecessary, and only yields a very modest voltage boost. 

When comparing Case (1c) and Case (3), we see that we pick up $\approx 2\%$ from the intermediate mirror in the case of 2 bandgaps, when assuming that the cells are optically thin to the internal luminescence. For more bandgaps, we pick up a greater absolute efficiency increase, with diminishing returns. When we make the comparison with the optically thin case rather than the optically thick case, we see a greater boost in efficiency with proper mirror design. As we assume step function absorption in this analysis, making the assumption that the cell is optically thin to luminescence appears to be contradictory. In a real material, the absorption of the luminescence depends on the degree of overlap between the luminescence spectrum and absorption spectrum. As the absorption is usually very weak below the band-edge, this is actually a reasonable approximation.

An intermediate mirror has the dual burden of reflecting the internally fluorescent photons and transmitting below bandgap photons. We thus propose an air gap with an antireflection coating to serve as the intermediate mirror, using angular selectivity by total internal reflection to achieve frequency selectivity. Though the air gap presents manufacturing difficulties, it is a feasible architecture, as demonstrated experimentally in \cite{sheng_device}.

\section*{Acknowledgements}

We acknowledge Dr. Myles Steiner for helpful discussions regarding the difference between the optically thin and thick cases. V.G. was supported as part of the DOE ``Light-Material Interactions in Energy Conversion" Energy Frontier Research Center under grant DE-SC0001293; C.H. was supported by the U.S. Department of Energy under Contract No. DE-AC36-08-GO28308 with the National Renewable Energy Laboratory.

\bibliographystyle{ieeetr}
\bibliography{manuscript}

\end{document}